\documentclass[10pt,journal,compsoc]{IEEEtran}
\usepackage{amsmath,amsfonts}
\usepackage{algorithmic}
\usepackage{algorithm}
\usepackage{array}
\usepackage{subfigure}
\usepackage{textcomp}
\usepackage{stfloats}
\usepackage{url}
\usepackage{verbatim}
\usepackage{graphicx}
\usepackage{cite}
\usepackage{booktabs}
\usepackage{color}
\usepackage{bm}
\usepackage{multirow}
\usepackage[colorlinks, linkcolor=red, anchorcolor=red, citecolor=blue]{hyperref}
\usepackage{pifont}
\usepackage{enumitem}
\begin{document}

\title{Event-Enhanced Snapshot Compressive Videography at 10K FPS}

\author{Bo~Zhang, Jinli~Suo and Qionghai~Dai
\thanks{B. Zhang, J. Suo and Q. Dai are with the Department of Automation, Tsinghua University, Beijing 100084, China. J. Suo is also with the Institute for Brain and Cognitive Sciences, Tsinghua University, Beijing 100084, China, and Shanghai Artificial Intelligence Laboratory, Shanghai 200232, China.\\E-mail: jlsuo@tsinghua.edu.cn.}
\thanks{Corresponding author: Jinli Suo.}
}


\IEEEpubid{0000--0000/00\$00.00~\copyright~2021 IEEE}

\IEEEtitleabstractindextext{%

\begin{abstract}
Video snapshot compressive imaging (SCI) encodes the target dynamic scene compactly into a snapshot and reconstructs its high-speed frame sequence afterward, greatly reducing the required data footprint and transmission bandwidth as well as enabling high-speed imaging with a low frame rate intensity camera. 
In implementation, high-speed dynamics are encoded via temporally varying patterns, and only frames at corresponding temporal intervals can be reconstructed, while the dynamics occurring between consecutive frames are lost. 
To unlock the potential of conventional snapshot compressive videography, we propose a novel hybrid ``intensity$+$event” imaging scheme by incorporating an event camera into a video SCI setup. 
Our proposed system consists of a dual-path optical setup to record the coded intensity measurement and intermediate event signals simultaneously, which is compact and photon-efficient by collecting the half photons discarded in conventional video SCI. Correspondingly, we developed a dual-branch Transformer utilizing the reciprocal relationship between two data modes to decode dense video frames.
Extensive experiments on both simulated and real-captured data demonstrate our superiority to state-of-the-art video SCI and video frame interpolation (VFI) methods. Benefiting from the new hybrid design leveraging both intrinsic redundancy in videos and the unique feature of event cameras, we achieve high-quality videography at 0.1ms time intervals 
with a low-cost CMOS image sensor working at 24 FPS.

\end{abstract}

}

\maketitle

\begin{figure*}[t]
    \centering
    \includegraphics[width=1\linewidth]{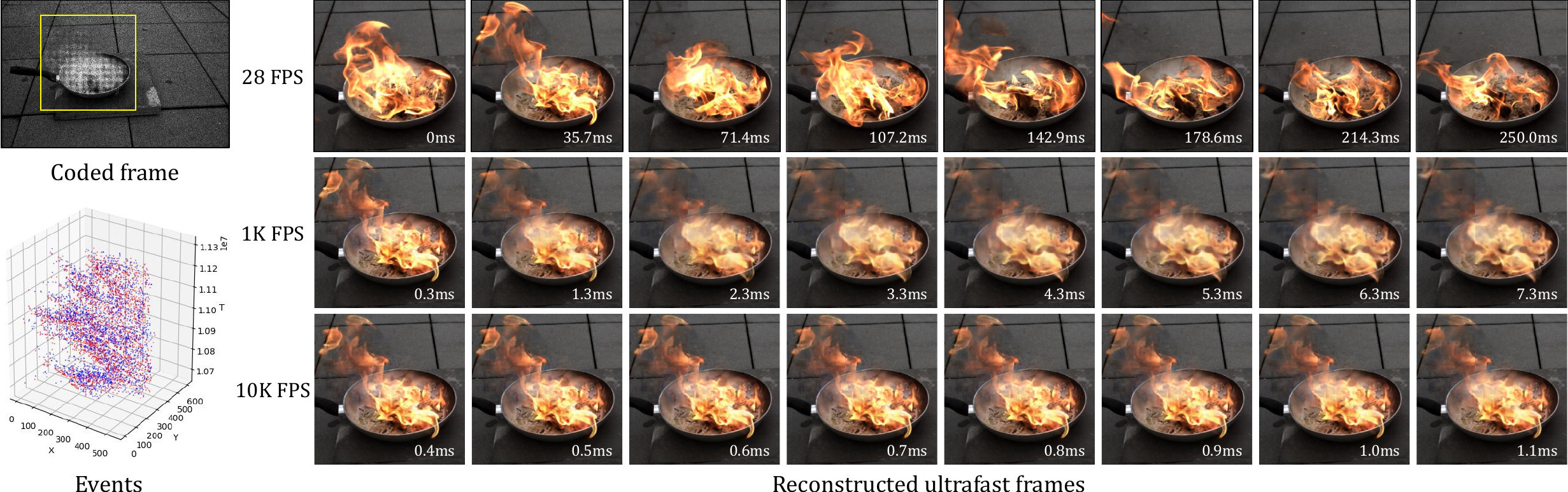}
    \caption{One representative example demonstrating the inputs and outputs of the proposed imaging scheme. From a coded snapshot and a series of events happening within the exposure elapse, we can reconstruct around 2500 frames (corresponding to 10K frames per second) recording the dynamics of the target scene with high fidelity.}
    \label{fig:teaser}
    \vspace{-1mm}
\end{figure*}

\section{Introduction}

\IEEEPARstart{H}{igh-speed} photography has been a long-sought goal in the realm of computer vision, as numerous vision-related applications merit a high imaging speed, such as TV broadcasting, autonomous driving, factory automation inspection, etc. However, high-speed sensor chips are mainly for scientific purposes and too expensive for commercial use. Some smartphones can support high-speed imaging up to 1K FPS but the duration is limited to only several seconds due to demanding memory and storage consumption. 

The research of high-speed imaging has continued for decades and achieved impressive progress. Over the past years, data-driven approaches have been flourishing and researchers have proposed many video frame interpolation approaches to produce dense video frames from low-speed ones computationally by predicting motions between boundary frames. Although temporal interpolation is a highly underdetermined problem, impressive results have been achieved. Especially, recent methods integrate novel neuromorphic sensors~\cite{Tulyakov_2021_CVPR}\cite{Tulyakov21cvpr} to provide additional dynamic information, which promises better interpolation performance and may reconstruct video frames up to 1K FPS. In general, these purely data-driven methods usually produce decent results in the cases of gradual motion but often hallucinate discontinuous motion details in the middle frames for intense or drastic dynamics.

Alternatively, computational photography designs new imaging devices for encoded scene recording and develops corresponding reconstruction algorithms to decode the target scene, which incorporates physical models both optically and computationally to raise the imaging capability, including high dynamics. 
As a representative high-speed imaging scheme, video snapshot compressive imaging captures a coded snapshot using a conventional low-speed intensity camera and reconstructs high-speed frames with specific algorithms. This imaging scheme relaxes the requirement of transmission bandwidth and storage memory, and leverages benefits from both optics and algorithms. 
Video SCI has experienced tremendous advances in both setups and reconstruction algorithms. 
With the rapid development of deep neural networks, 
video SCI reconstruction algorithms using deep networks~\cite{Wang_2021_CVPR,10.1007/978-3-030-58586-0_16,9965744} have achieved promising performance. 
The maturity of both imaging setups and algorithms enables reconstructing up to 50 high-speed frames from a single snapshot~\cite{9965744}. 
However, a highly compressed measurement imposes big challenges on the visual reconstruction results of large and fast motions in complex scenes. 
Besides, the algorithms can only reconstruct frames matching the switch of coding masks, which means the information between the intermediate frames is lost. 

To unlock the upper limits of high-speed imaging at a small data footprint and low transmission bandwidth, we propose a novel imaging scheme to improve the frame rate and quality by combining physics-driven 
video SCI and data-driven dynamics-aware VFI,  leveraging the benefits from both regimes. 
The proposed imaging scheme includes a dual-path intensity-event optical setup to simultaneously capture encoded intensity snapshots at a certain compression rate and events containing intermediate information between frames, and a dual-branch Transformer to reconstruct and interpolate ultrafast video frames. 
The imaging setup bears the following three-fold advantages: firstly,  our design is photon-efficient, since an event camera is added to sense half count of photons that are otherwise blocked off in the original video SCI device; secondly, the optical setup is compact, easy to build and deploy; thirdly, the optical design enables imaging with two well-aligned FOVs, free from parallax in a stereo setup.
For the reconstruction and interpolation algorithm, we propose a dual-branch Transformer, consisting of a main branch that fuses both information modes (intensity and event) to reconstruct video SCI encoded frames and a sub-branch for interpolating desired-timestamp video frames. 
This design effectively aggregates information from two sources and efficiently reconstructs ultra-dense intermediate frames at a low computational cost. 
In addition, the frame at a given timestamp can be interpolated only using the sub-branch network from the pre-saved encoding features, 
thus greatly saving the inference time.

The proposed optical setup and the algorithm together constitute the main contributions of the paper, which are summarized as follows:
\begin{itemize}[leftmargin=*] 
    \item We develop a dual-path intensity-event optical setup to capture a well-aligned coded snapshot and event sequences simultaneously, corresponding to the encoded imposition of discrete video frames and intermediate dense dynamics respectively. The setup is compact and photon-efficient after a minimum adaption to a conventional video SCI setup. 
    \item We propose a dual-branch Transformer to learn a unified representation of the coded snapshot and event sequence, leveraging their both advantages to reconstruct dense frames at high fidelity.  
    \item We evaluate our method and compare it with state-of-the-art video SCI and VFI methods extensively on both simulated and real-captured intensity-event datasets, which 
    validates the feasibility and superiority of our method.
    \item We capture real highly dynamic scenes using our self-established setup and reconstruct ultrafast video frames with the proposed algorithm. Our approach demonstrates a high-throughput continuous ultrafast imaging capability with 5M pixel resolution working at 10K FPS, far beyond conventional high-speed photography.
\end{itemize}

In the following sections, we review the related work of video SCI, VFI, and event camera in Section~\ref{sec:relatedwork}. Then, we introduce the proposed imaging scheme from hardware and software aspects in Section~\ref{sec:method}. In Section~\ref{sec:experiments}, we conduct extensive experiments and analysis on several existing and self-captured datasets. Lastly in Section~\ref{sec:conclusion} we conclude the paper with a summary and discussions on future work.

\section{Related work}\label{sec:relatedwork}
\subsection{Video snapshot compressive imaging}
\subsubsection{Mathematical formulation}
Video snapshot compressive imaging can be deemed as an application of compressed sensing theory~\cite{Donoho06_CS}\cite{Candes06_Robust} in the field of optics and vision. The central process is to modulate high-speed frames with random varying masks, capture a coded snapshot with a low-speed camera, and reconstruct the high-speed ones afterward. Suppose $B$ high-speed frames 
$\lbrace\bm{I}_b\rbrace_{b=1}^B\in\mathbb{R}^{n_x\times n_y}$
are modulated by masks $\lbrace\bm{C}_b\rbrace_{b=1}^B\in\mathbb{R}^{n_x\times n_y}$ correspondingly, the snapshot measurement $\bm{Y}\in\mathbb{R}^{n_x\times n_y}$ is given by
\begin{equation}
\bm{Y}=\textstyle  \sum_{b=1}^{B}\bm{I}_b\odot C_{b}+\bm{G},
\label{eq:compr_imaging}
\end{equation}
where $\odot$ denotes Hadamard (element-wise) product and $\bm{G}$ denotes noise. 
We define $\bm{s}=[\bm{s}_1^\top,\dots,\bm{s}_B^\top]$, where $\bm{s}_{b}=$ vec$(\bm{I}_b)$, and let $\bm{D}_{b}=$ diag(vec($\bm{C}_b$)), for $b=1,\dots, B$, where vec$(\cdot)$ vectorizes the matrix by stacking the columns and diag$(\cdot)$ places the ensued vector into the diagonal elements of a diagonal matrix. The above operations provide the vector formulation of the sensing process of video SCI
\begin{equation}
\textstyle  \bm{y}=\bm{\Phi}\bm{s}+\bm{g},
\end{equation}
where $\bm{\Phi}=[\bm{D}_1,\dots,\bm{D}_B],\in\mathbb{R}^{n\times nB}$ is the sensing matrix with $n=n_{x}n_{y}$ and $\lbrace \bm{D}_b\rbrace_{b=1}^B$ being diagonal matrices, $\bm{s}\in\mathbb{R}^{nB}$ is the desired signal, and $\bm{g}\in\mathbb{R}^{n}$ denotes the vectorized noise component.

\subsubsection{Optical designs}
The key optical element of a video SCI system is the modulation masks that must vary fast enough to encode high-speed frames. The optical setups have experienced large progress in the decade. At first, the modulation patterns are realized by a moving lithography mask driven by a piezoelectric stage~\cite{Llull:13}. Further, various programmable spatial light modulators (SLM) such as DMD~\cite{8863920}\cite{qiao2020deep}, LCoS~\cite{6126254}\cite{5995542} are used to generate random binary masks with a faster modulation speed. The use of these modulation devices makes the overall optical path compact and flexible. For example, Zhang et al.~\cite{Zhang:22} combine diffractive optics with video SCI to enhance spatial and temporal resolution simultaneously. Lu et al.~\cite{lu2021DualviewSnapshot} propose a dual-view video SCI setup and an optical flow-aided recurrent neural network to achieve joint FOV and temporal compressive sensing. These systemic innovations on optical setups and corresponding algorithms largely alleviate the burden of capturing high throughput dynamic data and offer noticeable advantages of low-bandwidth, low-power and low-cost over pure algorithmic design. Therefore, exploration of video SCI systems is highly promising and valuable.

\subsubsection{Reconstruction algorithms}

In the past decade, reconstruction algorithms have been iteratively developed and big progress has been made, especially since deep learning was introduced to this field. They have experienced large improvements in terms of quality,  computational cost, and compression ratio. These algorithms can roughly be categorized as optimization-based, deep-learning-based, or a combination of both. Optimization-based approaches take sparse priors such as total variation (TV) constraint~\cite{Yuan16ICIP_GAP}, non-local self-similarity~\cite{dong2014compressive}\cite{maggioni2012video} and low-rank prior~\cite{liu2018rank} into optimization frameworks like generalized alternating projection (GAP)~\cite{liao2014generalized} and alternating direction method of multiplier (ADMM)~\cite{boyd2011distributed}. These approaches are highly flexible for quick adaption to different setups but are quite computation-consuming as they require iterative optimization to gradually obtain the final results.
With the advent of deep neural networks that features fast inference and high perormance, Plug-and-Play (PnP)~\cite{yuan2020plug}\cite{wu2023adaptive} approaches have been proposed to integrate a pre-trained network into the optimization framework to replace the time-consuming module and achieve faster and better-quality reconstruction. Along with the development of novel network architectures and frameworks, end-to-end deep networks for video SCI reconstruction have been devised elaborately, such as U-Net~\cite{qiao2020DeepLearning}, Unfolding Net~\cite{meng2020gap}, BIRNAT~\cite{10.1007/978-3-030-58586-0_16}, MetaSCI~\cite{Wang_2021_CVPR}, RevSCI~\cite{Cheng2021_CVPR_ReverSCI}. Recently, Transformer~\cite{NIPS2017_3f5ee243} has been proven feasible and effective in computer vision tasks as a new structure. Wang et al.~\cite{9965744} proposed STFormer, a Transformer-based network considering spatial and temporal attention, and achieves state-of-the-art performance for video SCI reconstruction, demonstrating the power of attention mechanism for sequential information modeling. In general, deep-learning solutions in video SCI exhibit satisfactory performance, robustness and generalization ability, and faster inference compared to conventional optimization frameworks, and are still in rapid development. 

\subsection{Event cameras}
Event cameras are bio-inspired neuromorphic sensors that response to pixel-wise logarithmic brightness changes $L(\bm{q}_k,t_k)\doteq\log(\bm{q}_k,t_k)$ asynchronously instead of conventional intensity frames. Specifically, at pixel $\bm{q}_k=(x_k,y_k)^T$ and time $t_k$, an event occurs when the brightness change since the last event at this location reaches a threshold $\pm T (T>0)$
\begin{equation}\label{eq:formation}
|\textstyle L(\bm{q}_k,t_k)-L(\bm{q}_k,t_k-\Delta t_k)|\geq T,
\end{equation}
where $\Delta t_k$ is the time elapsed since the last event at $\bm{q}$. An event sequence can be represented as a set of quadruples $\varepsilon(t_N)=\lbrace \mathbf e_k\rbrace_{k=1}^N=\lbrace(t_k,x_k,y_k,p_k)\rbrace_{k=1}^N$ with microsecond resolution where $p_k$ is a binary value (1 or -1) indicating the sign of the brightness change. Many event representations have been proposed, such as HFirst~\cite{7010933}, event frame~\cite{Rebecq_2017_BMVC}, event histograms~\cite{Sironi_2018_CVPR}, event-based time surfaces~\cite{7508476}, event spike tensor~\cite{Gehrig_2019_ICCV} and event volume~\cite{Zhu_2019_CVPR}, event voxel~\cite{Zhu_2018_ECCV_Workshops} etc.
With its unique working principle, an event camera features fast response speed, high dynamic range, and low latency. Event cameras have been successfully employed in high-speed and high-dynamic-range imaging~\cite{10.1007/978-3-030-58523-5_39,Chang_2023_CVPR}, video deblurring and frame interpolation~\cite{Tulyakov_2021_CVPR}\cite{9962797}\cite{10.1007/978-3-030-58598-3_41}, image reconstruction and super-resolution~\cite{Wang_2020_CVPR,I._2020_CVPR,Han_2021_ICCV} etc. In this paper, we utilize the advantages of event cameras to raise the throughput of video SCI and help render continuous dynamics further.

\subsection{Video frame interpolation}
Video frame interpolation aims to interpolate intermediate frame(s) from boundary frames, thus recovering a high-speed video sequence from the low-frame-rate counterpart. Most conventional approaches use boundary frames as input to predict middle frames via optical flow~\cite{baker2011database,10.1007/978-3-642-23094-3_20,6419791,10.1007/978-3-031-19781-9_36,Niklaus_2018_CVPR,Bao_2019_CVPR,Sim_2021_ICCV,Lu_2022_CVPR}, phase~\cite{Meyer_2015_CVPR}\cite{Meyer_2018_CVPR}, adaptive kernels~\cite{Niklaus_2017_CVPR,Niklaus_2017_ICCV,Lee_2020_CVPR} or direct synthesis~\cite{Kalluri_2023_WACV}\cite{Shi_2022_CVPR}. Despite the diversity of motion estimation strategies, these approaches only rely on boundary intensity frames during inference, which may lead to inaccurate motion estimations and unsatisfactory results in scenarios with ultrafast nonrigid motions. Low-frame-rate intensity frames fail to provide intermediate motion information. As a novel neuromorphic sensor, event cameras offer additional dynamic information due to their unique working principle and characteristics such as fast response and low latency and thus can be readily deployed for VFI as a supplement sensor~\cite{9962797}\cite{10.1007/978-3-030-58598-3_41}. Many methods have been proposed to aggregate the information in intensity frames and events for better interpolation performance such as Time Lens series~\cite{Tulyakov_2021_CVPR}\cite{Tulyakov21cvpr} and achieve promising visual results of interpolated frames at near 1K FPS compared to intensity-only approaches.

Indeed, VFI approaches can interpolate middle frames from boundary ones, thus realizing high-speed imaging by proper motion modeling, designing deep network architectures or incorporating additional dynamic information. 
The low-speed intensity camera used in these methods acts as a direct temporal downsampling of the scene dynamics and neglects the intrinsic redundancy within the recordings, so is less efficient in capturing fast-changing scenes.  
To further push the potential of high-speed imaging with low-bandwidth cameras, we propose a hybrid imaging scheme incorporating both SCI setups and interpolation by designing a dual-arm optical setup to encode higher dynamics and a dual-branch Transformer network to reconstruct dense video frames as well as interpolate dense-timestamp dynamics with high quality. 


\section{Method}\label{sec:method}
We propose a novel intensity-event snapshot compressive scheme for ultrafast imaging. The mathematical formulation, dual-path intensity-event optical setup for data capturing and dual-branch Transformer for reconstruction and interpolation will be introduced respectively in this section.

\subsection{Formulation of the dual-path intensity+event SCI scheme}
When capturing a scene with our dual-mode imaging scheme, the recorded events are denoted as $\bm{E}=\{\mathbf e_k\}=\{(t_k,x_k,y_k,p_k)\}$. Since the temporal resolution of these events is at millisecond-level, we can easily partition the event set $\bm{E}$ by timestamp. For example, one can split the events by frame duration, i.e. $\bm{E}=\{E_{1,2},E_{2,3},\dots,E_{B-1,B}\}$, where $E_{b,b+1}$ represents the events occurring between frames $b$ and $b+1$. Also, the events can be split such as $E_{b,b+0.5}$ for precise manipulation in frame interpolation.
Then, the target of our scheme is to reconstruct an intensity frame $\bm{I}_{t_d}$ given a desired timestamp $t_d$ with $\bm{Y}$ in Eq.~\eqref{eq:compr_imaging} and $\bm{E}$ as input:
\begin{equation}\label{eq:formulation}
\bm{I}_{t_d}=\mathcal{F}(\bm{Y}, \bm{E}, t_d, \theta),
\end{equation}
where $\mathcal{F}$ denotes the reconstruction algorithm which is implemented as a deep neural network in our case and $\theta$ represents its parameters.

\subsection{Dual-path intensity+event SCI setup}

\begin{figure}[t]
    \centering
    \begin{minipage}[ht]{0.8\linewidth}
    \subfigure[]
    {\includegraphics[width=0.98\linewidth]{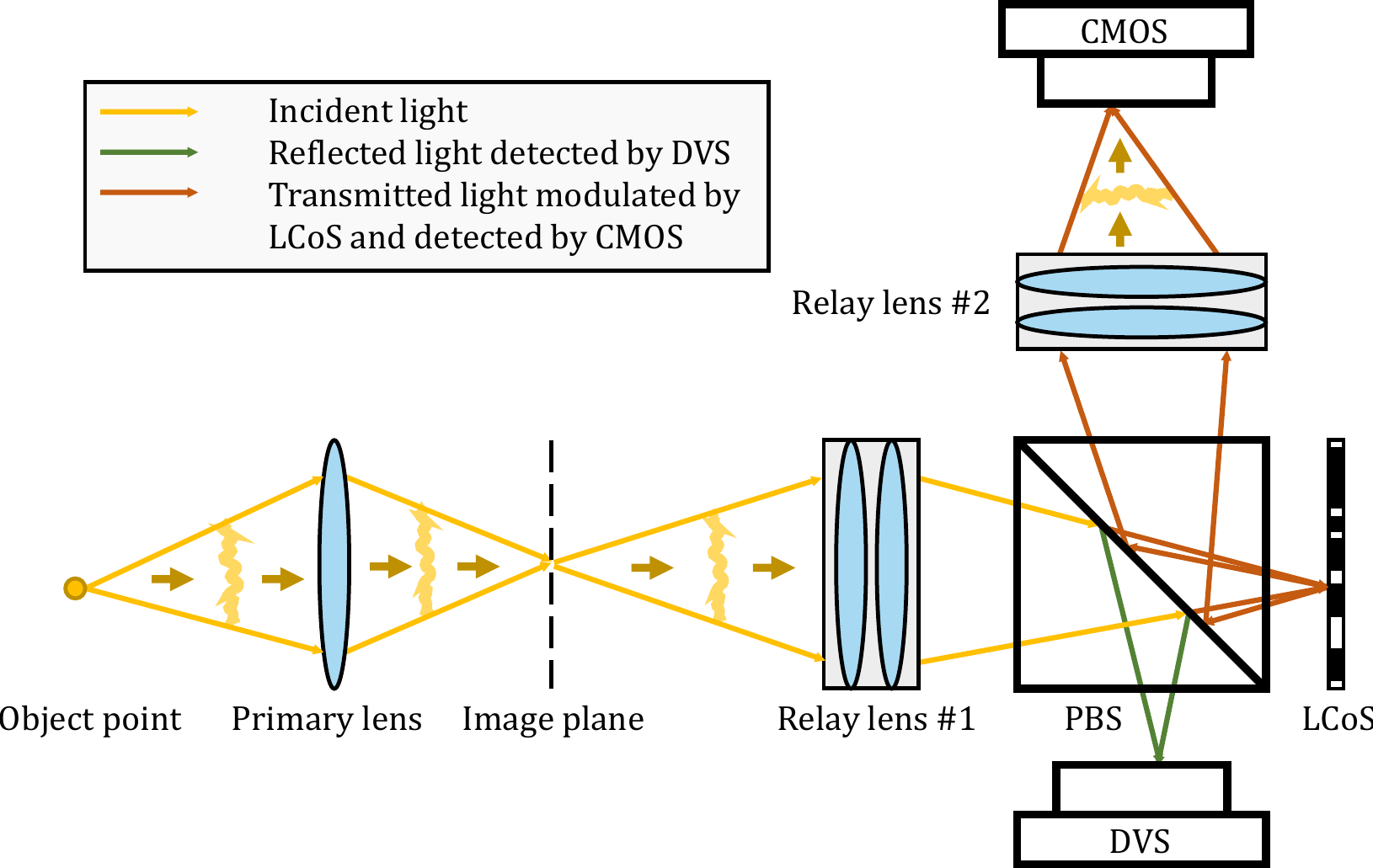}}
    \end{minipage}\\
    \vspace{-1mm}
    \begin{minipage}[ht]{0.8\linewidth}
    \subfigure[]
    {\includegraphics[width=0.98\linewidth]{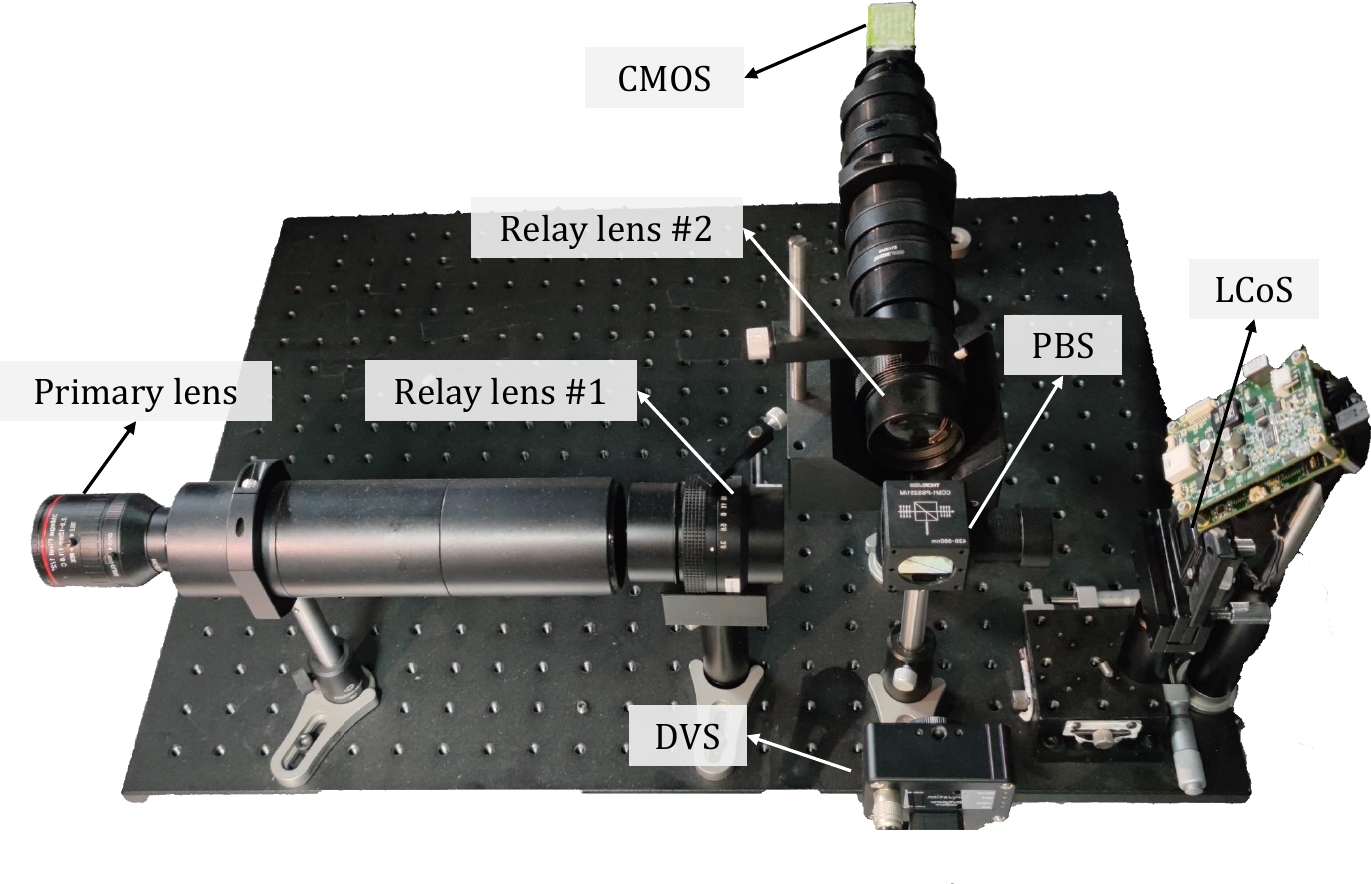}}
    \end{minipage}
    \vspace{-2mm}
    \caption{The proposed dual-arm imaging setup. (a) and (b) display the schematic optical path and established prototype respectively. The incident light of a scene point converges at the image plane, which is transferred to the 2nd image plane by Relay lens $\#1$ and the 3rd plane by Relay lens $\#2$. Before arriving at the 2nd imaging plane,  the light beam is split by a PBS into two halves, one detected with a DVS and the other modulated by LCoS and then captured with a CMOS.}
    \label{fig:optics}
    \vspace{-2mm}
\end{figure}

We develop a compact dual-arm optical system to simultaneously capture the coded intensity measurement at fixed compression ratio and events sensing the intermediate dynamic information. The overview and prototype of the setup are shown in Fig.~\ref{fig:optics} (a) and (b) respectively. Our system consists of a primary lens (KOWA LM50H, f=50 mm), a pair of high-quality large-field relay lenses (Chiopt, LS1610A) to transfer the first image plane to the sensor and extend the working distance for light modulation, and a beam splitter partitioning the incoming light into two branches for separate sensing. In one branch, an LCoS (ForthDD, QXGA-3DM, $2048\times1536$ pixels, 4.5k refresh rate) is placed at the second image plane to randomly modulate the high-speed frames and the modulated signals pass a second relay lens to form snapshot coded measurement which is detected with a low-speed CMOS camera (JAI, GO-5100C-USB, $2056\times 2464$ pixels). In the other branch, the unmodulated optical signals are sensed and converted to event signals via a DVS (iniVation, DVXplorer, $480\times 640$ pixels). Other optical components include a PBS (Thorlabs, CCM1-PBS251/M).
In our setup, we use a signal generator to generate pulses that synchronize LCoS, CMOS, and DVS. When the signal generator sends a pulse, the LCoS starts displaying varying patterns, the CMOS starts exposure and recording, and the DVS records a special event with precise timestamp in the output event sequence. This timestamp allows precise temporal alignment between intensity measurement and events.

Compared to conventional intensity-only SCI setup, our optical setup is photon-efficient, since half of the photons are otherwise discarded by the PBS. Alternatively, the photons containing dynamic information of the scene are recorded with a DVS, which is further utilized in our algorithm to enhance the reconstruction and interpolation performance. Additionally, our design requires only slight adaption to the conventional setup by conveniently adding an extra DVS, therefore it can be quickly built on an LCoS-based SCI setup.

\begin{figure*}[]
    \centering
    \includegraphics[width=\linewidth]{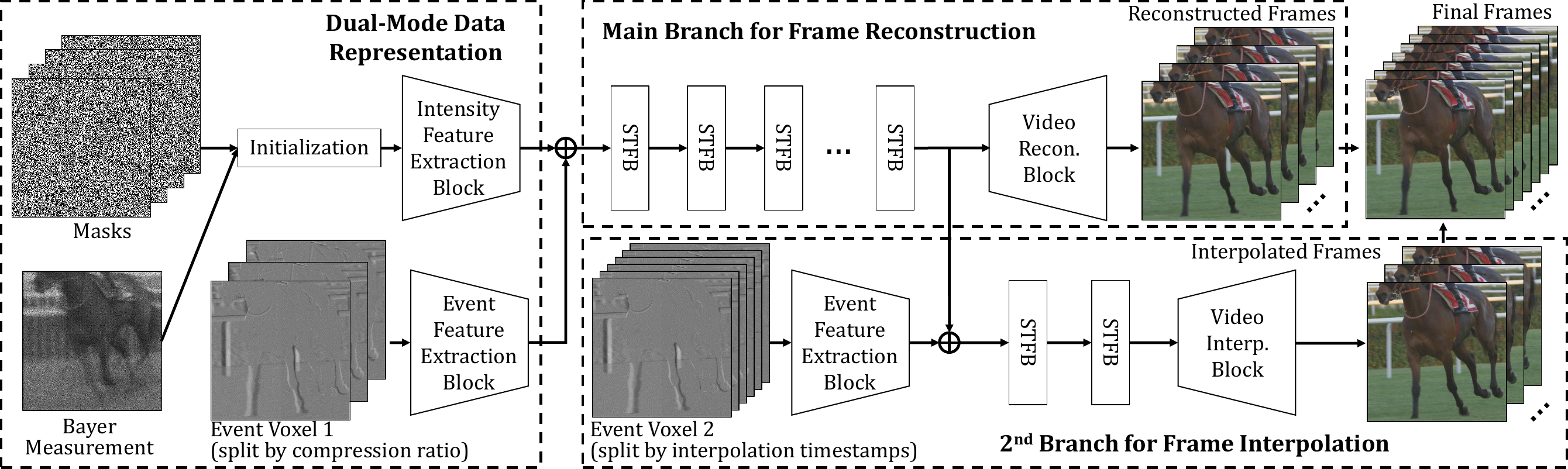}
    \caption{The network structure of the proposed dual-branch Transformer for dense video reconstruction. The dual-mode data representation module provides initialization of intensity frames and generates intensity as well as event tokens. The main branch processes two types of tokens and reconstructs frames at the encoding frame rate, and the 2nd branch additionally takes in timestamp-aware event tokens and produces much denser sequence with high fidelity.}
    \label{fig:network}
\end{figure*}

\subsection{Dual-branch Transformer reconstructing highly dynamic scenes}
With both intensity measurement and intermediate events acquired, an algorithm consuming the dual-mode data is demanded to effectively and efficiently reconstruct and interpolate ultra-high-speed video frames. To this end, we propose a dual-branch Transformer with the overview shown in Fig.~\ref{fig:network}. The deep neural network is composed of a dual-mode data representation module, which pre-processes intensity measurement and generates intensity and event tokens, and two branches for high-speed frame reconstruction and interpolation, i.e. one main branch for information integration as well as dense frame reconstruction and a sub-branch for dense-timestamp frame interpolation.

The framework starts from a preprocessing step, aggregating information from both the coded intensity measurement and intermediate events. 
Firstly, the calibrated masks and captured intensity measurement are used to obtain an initial estimation of the modulated frames, which are processed by the intensity feature extraction block to generate intensity tokens. 
At the same time, we convert the events into voxel grid representation $\bm{V}=\{V_{1,2},V_{2,3},\dots,V_{B-1,B}\}$ following the procedure described in~\cite{Zhu_2018_ECCV_Workshops}. The events are sliced at corresponding frame intervals, so each event voxel contains the dynamics between two boundary frames. These voxels are fed into the event feature extraction block for information embedding and to generate event tokens. 

In the first branch, two groups of tokens are aggregated to form the final tokens which are fed into the proceeding spatial-temporal Transformer blocks (STFB). At this point, the intensity and event information are well fused. 
Further, spatial and temporal correlation is built through a chain of STFBs, and the processed tokens are passed to the video reconstruction block to reconstruct the dense frames that are encoded by the intensity camera. The achieved frame rate is determined by the compression ratio $B$ of the system.

The learned tokens containing all information serve as the input for the second branch. The events are again converted into event voxels but with finer partitioning. Suppose we aim to interpolate the desired frame $\bm{I}_{1+f},\bm{I}_{2+f},\dots,\bm{I}_{B+f}$, where $f$ is a fraction between 0 and 1 indicating the interpolation position, and then we need to convert event voxel $\bm{V}$ into $\bm{V_1}=\{V_{1,1+f},V_{2,2+f},\dots,V_{B,B+f}\}$ and $\bm{V_2}=\{V_{1+f,2},V_{2+f,3},\dots,V_{B-1+f,B}\}$. We stack $\bm{V_1}$ and $\bm{V_2}$ and feed the result into an event feature extraction block to obtain the interpolation control tokens, which are aggregated with the reserved tokens in the main branch. Then, the fused tokens are passed through two additional STFBs and processed with a video interpolation block to interpolate frames at the desired timestamps. This practice utilizes the reserved tokens with readily built spatial and temporal correlation, thus saving computation during interpolation.

Both intensity/event feature extraction blocks are composed of 3D convolution layers and LeakyRELU~\cite{xu2015empirical} to learn spatial and temporal correlation within the tokens. Similarly, video reconstruction/interpolation blocks consist of 3D transpose convolution layers to reconstruct video frames from tokens. STFB is composed of three parts: spatial self-attention (SSA) branch, temporal self-attention (TSA) branch, and grouping resnet feed-forward (GRFF) network. The details of the initialization block, feature extraction block, video reconstruction/interpolation blocks, and STFB can be referred to in~\cite{9965744}.

\section{Experiments}\label{sec:experiments}
In this section, we experimentally validate the proposed imaging scheme by comparison with existing low-bandwidth high-speed imaging strategies, i.e. computational imaging scheme---video SCI, pure-computational technique---VFI as well as the combination of both. 
In terms of quantitative comparison, we use peak signal-to-noise-ratio (PSNR) and structural similarity (SSIM)~\cite{1284395} to evaluate the performance of different methods on simulation datasets.
Regarding visual comparison, we demonstrate our superiority in capturing fast dynamics on multiple simulation and real datasets captured with our self-established setup. 

\begin{figure*}[t]
    \centering  \includegraphics[width=\linewidth]{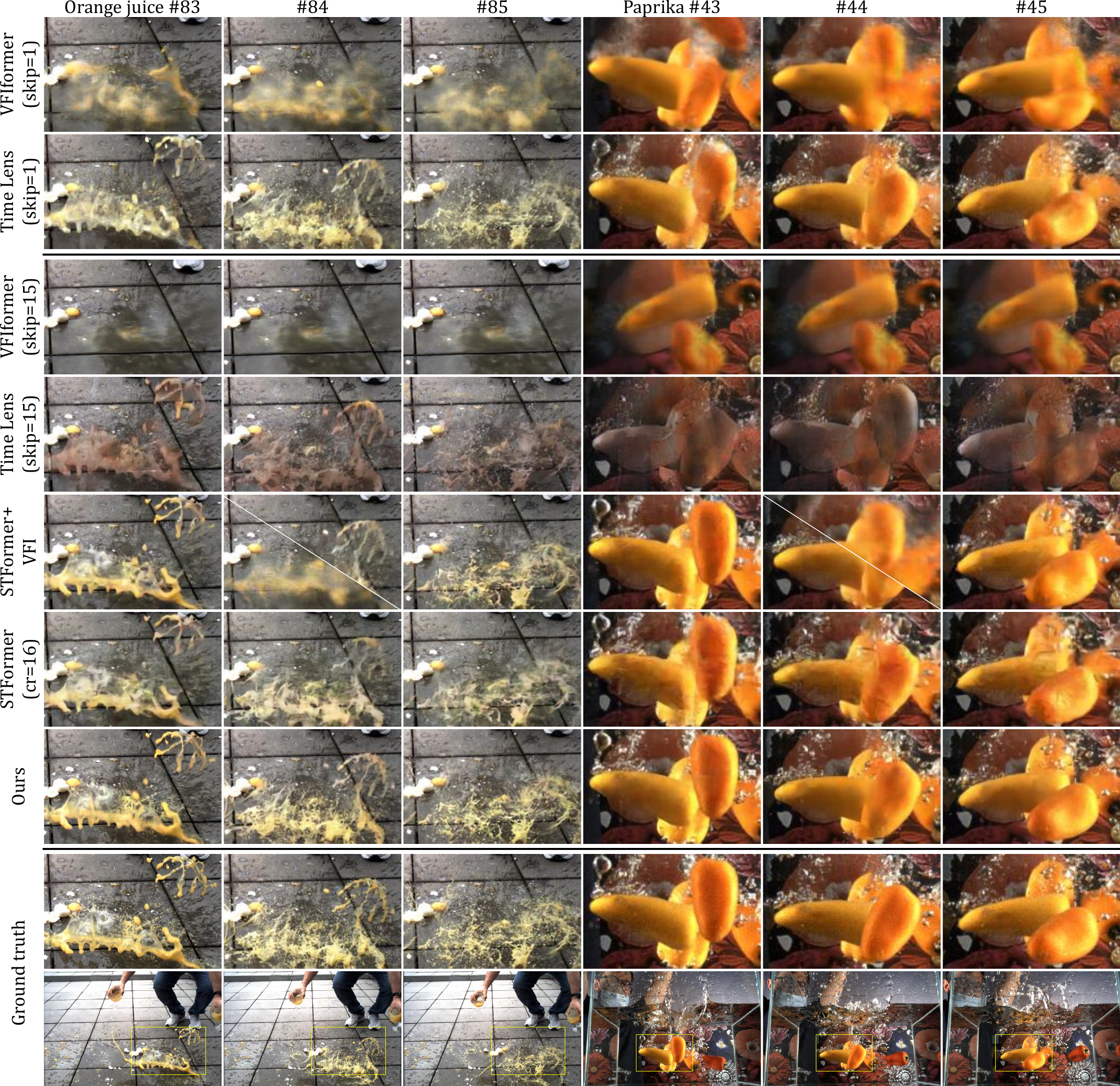}
    \caption{Reconstructed consecutive frames of two challenging scenes from the BS-ERGB test set. VFIformer produces blurry results for such fast motions even with only skip=1 and fails with skip=15. Time Lens predicts better motion with auxiliary events but suffers color distortion with a larger skip. STFormer+VFI (lower left: VFIformer, upper right: Time Lens) reconstructs frames with decent quality at the frames corresponding to the encoding device but interpolates blurry results at the intermediate timestamps. In comparison, our method reconstructs high-quality frames with consistently accurate dynamics across dense timestamps.}
    \label{fig:res_bsergb}
\end{figure*}

\begin{table*}[t]
    \centering 
    \vspace{3mm}
    \caption{Average PSNR (dB) and SSIM of reconstruction results by the proposed approach and other baseline schemes on the BS-ERGB test set containing real-captured aligned color video frames and events. 
    For our method, 16 frames are reconstructed from a snapshot encoding 8 intensity frames and the intermediate events. For STFormer+VFI approaches, 8 frames are reconstructed from a snapshot and then middle frames are interpolated. The best scores are highlighted in \textbf{bold}.}
    \begin{center}
    \begin{tabular}{l|ccccc|cc}
        \midrule
        & \quad Video SCI?\quad  &\quad\quad VFI? \quad\quad& Use frames? & Use events? & Frame sampling ratio & \quad\quad PSNR \quad\quad&\quad SSIM \quad\quad\\
        \toprule
        VFIformer (skip=1) & & \checkmark & \checkmark &&  50\% & 25.67 & 0.7825 \\
        Time Lens (skip=1) && \checkmark & \checkmark & \checkmark & 50\% & 27.16 & 0.7847 \\
        \midrule
        VFIformer (skip=15) & & \checkmark & \checkmark && 6.7\% & 19.86 & 0.6673 \\
        Time Lens (skip=15) && \checkmark & \checkmark & \checkmark & 6.7\% & 23.96 & 0.7358 \\
        STFormer+VFIformer & \checkmark & \checkmark & \checkmark && 6.7\% & 26.61 & 0.7704 \\
        STFormer+Time Lens & \checkmark & \checkmark & \checkmark & \checkmark & 6.7\% & 27.31 & 0.7748 \\
        STFormer (CR=16) & \checkmark && \checkmark && 6.7\% & 27.95 & 0.7801 \\
        \midrule
        Ours & \checkmark & \checkmark & \checkmark & \checkmark & 6.7\% & \textbf{29.06} & \textbf{0.8051} \\
        \midrule
    \end{tabular}
    \end{center}
    \label{tab:res_real}
\end{table*}

\begin{table*}[h]
    \centering    
    \vspace{1mm}
    \caption{Average PSNR (dB) and SSIM of frame reconstruction results of different imaging schemes on 6 color benchmark test sequences. Here the event signals are simulated using VID2E and reconstruction models are trained on the DAVIS 2017 dataset. The best scores are highlighted in \textbf{bold}.}
    \begin{center}
    \begin{tabular}{l|ccccccc}
        \toprule
        Dataset & Beauty & Bosphorus & Jockey & Runner & ShakeNDry & Traffic & Average \\
        \toprule
        VFIformer (skip=1) & 36.05 / 0.8912 & 42.42 / 0.9766 & 30.68 / 0.9044 & 39.99 / 0.9863 & 29.34 / 0.7665 & 26.72 / 0.8970 & 34.20 / 0.9037 \\
        Time Lens (skip=1) & 34.90 / 0.8626 & 40.50 / 0.9651 & 27.88 / 0.8472 & 38.05 / 0.9810 & 28.28 / 0.7210 & 27.00 / 0.9651 & 32.77 / 0.8769 \\
        \midrule
        VFIformer (skip=15) & 26.61 / 0.7879 & 32.61 / 0.9046 & 20.19 / 0.7171 & 27.29 / 0.9174 & 22.82 / 0.7707 & 17.02 / 0.6595 & 24.43 / 0.7812 \\
        Time Lens (skip=15) & 25.66 / 0.7311 & 27.37 / 0.7820 & 19.50 / 0.6691 & 24.05 / 0.8237 & 22.82 / 0.6636 & 16.04 / 0.5629 & 22.57 / 0.7054 \\
        STFormer+VFIformer & 35.90 / 0.8816 & 39.21 / 0.9585 & 34.89 / 0.9195 & 39.97 / 0.9750 & 32.13 / 0.8315 & 28.30 / 0.8963 & 35.07 / 0.9104 \\
        STFormer+Time Lens & 35.77 / 0.8785 & 39.12 / 0.9574 & 33.79 / 0.9050 & 39.16 / 0.9728 & 31.70 / 0.8179 & 28.38 / 0.8947 & 34.65 / 0.9044 \\
        STFormer (CR=16) & 35.74 / 0.8764 & 37.79 / 0.9504 & 35.42 / 0.9118 & 39.33 / 0.9696 & 32.49 / 0.8414 & 27.86 / 0.8853 & 34.77 / 0.9058 \\
        \midrule
        Ours  & \textbf{40.00} / \textbf{0.9575} & \textbf{43.63} / \textbf{0.9892} & \textbf{40.37} / \textbf{0.9733} & \textbf{46.40} / \textbf{0.9952} & \textbf{41.41} / \textbf{0.9808} & \textbf{40.63} / \textbf{0.9923} & \textbf{42.07} / \textbf{0.9814} \\
        \bottomrule
    \end{tabular}
    \end{center}
    \label{tab:res_color}
\end{table*}

\begin{table*}[h]
    \centering
    \vspace{2mm}
    \caption{Average PSNR (dB) and SSIM of frame reconstruction results of different imaging schemes on 4 gray-scale benchmark test sequences. Here the event signals are simulated using VID2E and models are trained on the DAVIS 2017 dataset. The best scores are highlighted in \textbf{bold}.}
    \begin{center}
    \begin{tabular}{l|ccccc|cc}
        \midrule
        Dataset & Aerial & Crash & Kobe & Traffic & Average & Params (M) & FLOPs (T) \\
        \toprule
        VFIformer (skip=1) & 36.02 / 0.9825 & 29.82 / 0.9671 & 27.44 / 0.9111 & 34.14 / 0.9806 & 32.12 / 0.9627 & 24.2 & 5.12  \\
        Time Lens (skip=1) & 32.20 / 0.9630 & 28.66 / 0.9555 & 26.25 / 0.8808 & 30.53 / 0.9652 & 29.41 / 0.9411 & 79.2 & 3.36 \\
        \midrule
        VFIformer (skip=15) & 24.41 / 0.8355 & 24.61 / 0.8930 & 17.94 / 0.5098 & 22.28 / 0.8638 & 22.31 / 0.7857 & 24.2 & 5.12 \\
        Time Lens (skip=15) & 21.00 / 0.7094 & 20.36 / 0.7478 & 17.70 / 0.5014 & 15.37 / 0.4467 & 18.61 / 0.6013 & 79.2 & 3.36 \\
        STFormer+VFIformer & 29.85 / 0.9302 & 29.83 / 0.9566 & 30.82 / 0.9295 & 30.58 / 0.9522 & 30.30 / 0.9437 & 43.7 & 5.53 \\
        STFormer+Time
        Lens & 29.44 / 0.9263 & 29.59 / 0.9529 & 30.31 / 0.9171 & 29.79 / 0.9460 & 29.79 / 0.9368 & 98.7 & 4.65 \\
        STFormer (CR=16) & 29.21 / 0.9197 & 29.48 / 0.9506 & 31.71 / 0.9348 & 29.07 / 0.9350 & 29.87 / 0.9350 & 19.5 & 5.95 \\
        \midrule
        Ours  & \textbf{44.98} / \textbf{0.9964} & \textbf{43.34} / \textbf{0.9942} & \textbf{44.24} / \textbf{0.9884} & \textbf{44.06} / \textbf{0.9970} & \textbf{44.16} / \textbf{0.9940} & 32.2 & 5.26 \\
        \midrule
    \end{tabular}
    \end{center}
    \label{tab:res_gray}
\end{table*}

\subsection{Simulation experiment}
\subsubsection{Datasets}
We used two datasets for network training: large-scale color video dataset---DAVIS 2017~\cite{Pont-Tuset_arXiv_2017} and real-captured intensity-event dataset---BS-ERGB~\cite{Tulyakov21cvpr}. 
For the model trained on DAVIS 2017, we use a video-to-event method~\cite{Gehrig_2020_CVPR} to generate simulated events from video frames and test on 
six classical benchmark color sequences, including \texttt{Beauty}, \texttt{Bosphorus}, \texttt{Jockey}, \texttt{Runner}, \texttt{ShakeNDry} and \texttt{Traffic} with a size of 512$\times$512$\times$3$\times$8 pixels. We also evaluate the model with gray-scale outputs on 4 benchmark videos, including \texttt{Aerial}, \texttt{Crash}, \texttt{Kobe}, and \texttt{Traffic} with a size of 256$\times$256$\times$8 pixels. 
In comparison, the BS-ERGB dataset contains aligned color video frames and intermediate events, so we directly evaluate the well-trained model on the test set. 


\subsubsection{Implementation details}
For network training, we use the pre-trained weights of a standard one-branch STFormer to initialize the main branch of our model and then additionally train for 20 epochs for DAVIS 2017. Further, we use the optimized model to train on BS-ERGB for 10 epochs to compensate for the domain gap between simulated events and real-captured ones. During training, we adopt data augmentation including random flipping and random cropping for BS-ERGB and additional random scaling for DAVIS 2017. The patch size for training is 128$\times$128 pixels and the initial learning rate is set to be 0.0001 using Adam optimizer~\cite{kingma2017adam}. All the experiments in this paper are implemented with PyTorch on NVIDIA RTX 3090 GPUs.

For sequence partitioning, we first split the video frames of each scene into 16-frame snippets indexed $\#1, \#2, ..., \#16$. Then, we encode 8 intensity frames at one interval 
for each snippet into one snapshot, i.e. $\#1, \#3, ..., \#15$, and reconstruct all 16 frames with the input snapshot measurement and events.

\subsubsection{Comparative methods}
We compare our method with state-of-the-art methods of video SCI and VFI and the combination of the two, including (i) VFIformer~\cite{Lu_2022_CVPR}, a frame-based interpolation method using a Transformer structure and achieving SOTA performance; (ii) Time Lens~\cite{Tulyakov_2021_CVPR}, an intensity-event approach for VFI which achieves impressive performance on real dataset; (iii) STFormer~\cite{9965744}, which reconstructs frames from snapshot coded measurement with spatial-temporal attention modules for Transformer and achieves SOTA performance for video SCI; (iv) Direct combination of STFormer and VFIformer, i.e.,  first reconstructing frames from snapshot measurement with STFormer and then interpolating additional frames with VFIformer; (v) Combination of STFormer and Time Lens, i.e, first reconstructing frames from snapshot measurement with STFormer and then interpolating additional frames with Time Lens. 

\newpage
For (i) and (ii), we evaluate their performance with the skip set to be 1 and 15 respectively. Since VFIformer is a frame-based VFI approach, we directly use the pre-trained model for evaluation. In comparison, we finetune the provided Time Lens model on two training sets for 20 epochs, since there exists a large domain gap between synthetic and real event data as mentioned in the original paper~\cite{Tulyakov_2021_CVPR}. For (iii), the compression ratio is set to be 16. For (iv) and (v), a CR=8 STFormer model reconstructs 8 frames from the input snapshot and then a VFI approach is implemented to interpolate middle frames from the reconstructed results. All these algorithms are trained or fine-tuned with their default settings. 


\begin{figure*}[h]
    \centering
    \includegraphics[width=1\linewidth]{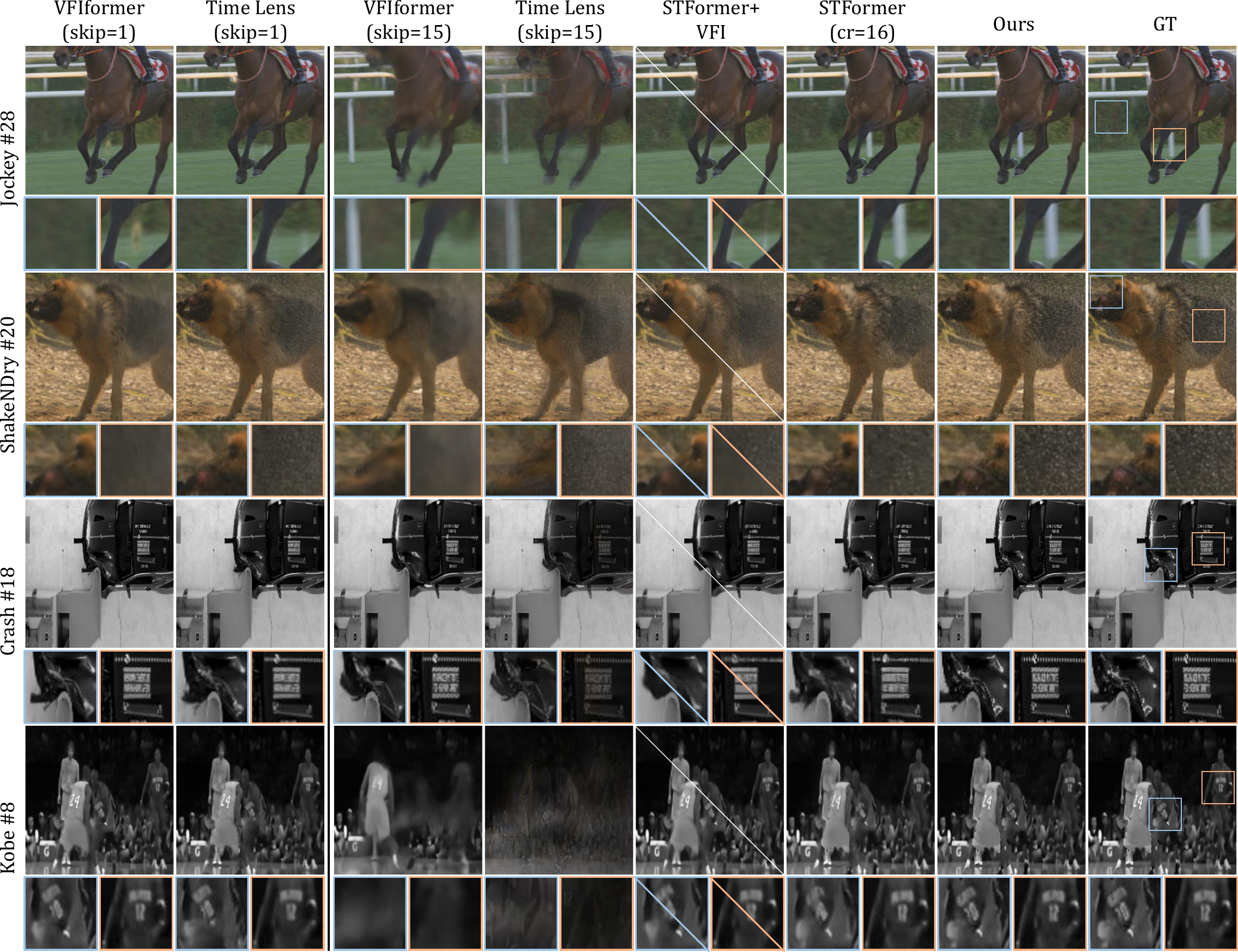}
    \caption{Visual results of exemplary frames in color and gray benchmark datasets, with zoomed-in views provided for a clearer performance comparison. 
    For the results by STFormer+VFI, the lower left corner is by VFIformer, and the upper right is by Time Lens.}
    \label{fig:res_color-gray}
\end{figure*}

\subsubsection{Quantitative results}
The performances of reconstruction and interpolation results in terms of PSNR and SSIM are summarized in Tables~\ref{tab:res_real}, \ref{tab:res_color} and \ref{tab:res_gray}. Our method consistently achieves the best results on four groups of test datasets among all the algorithms even with a 6.7\% frame sampling ratio. 

For the real-captured BS-ERGB dataset, our method obtains 29.06 dB and 0.8051 with an obvious advantage over the second-best scores by STFormer (CR=16). VFI approaches with skip=15 lead to much lower scores than the case of skip=1, which demonstrates the great challenge of this real dataset under a high sampling ratio. The combination of video SCI and VFI produces reasonable results comparable to STFormer (CR=16) largely because the video SCI approach is theory-based and the coded measurement contains high-speed information at fixed CR, leading to a decent performance. Time Lens achieves better results than VFIformer since it aggregates the dynamic information contained in events together with intensity frames for this real dataset with fast motions, fierce deformation, and large displacement such as object falling and smashing.

\newpage
On test sets with simulated events, our method shows more obvious superiority to the comparative methods. The overall trend is similar to that on the BS-ERGB dataset with one exception. Intensity-based VFIformer produces higher scores than intensity-event-based Time Lens. One possible explanation is that, even though the Time Lens model is finetuned on simulation datasets sufficiently, there exists a domain gap between the training and test sets, which may harm the performance. Another reason may be the motion in the test set is relatively small compared to the real BS-ERGB, thus reducing the advantage of auxiliary events.

\subsubsection{Visual results}
Some representative frames in three datasets are shown in Figs.~\ref{fig:res_bsergb} and~\ref{fig:res_color-gray} respectively. Fig.~\ref{fig:res_bsergb} shows the results of 3 consecutive frames of two scenes---\texttt{spilling juice} and \texttt{paprika dropping in water}---with very fast dynamics. VFIformer cannot handle such challenging scenes and produces blurry output frames for skip=1 and missing contents for skip=15. In comparison, the interpolated frames of Time Lens (skip=1) contain sharper details but still fail in the skip=15 scenario with color distortion and blur artifacts. STFormer+VFI generates good boundary frames but the quality of the middle frame is determined by the used VFI approach. STFormer (CR=16) produces visually consistent results, but the sharpness of dynamic parts and overall quality are inferior to our method. In general, our scheme achieves visually pleasant results with accurate motion prediction and temporal consistency.

We show the results of four representative scenes of classical color and gray simulation test sets in Fig.~\ref{fig:res_color-gray}. The performance is consistent with the scores in Tables~\ref{tab:res_color} and~\ref{tab:res_gray}. Generally, VFI-based methods produce better results with smaller skips, but some fast motions still cannot be reconstructed accurately either with or without auxiliary events. For example,
the existence and position of the white railing are incorrectly predicted in \texttt{Jockey}, and the interpolated splashing drops are blurred in \texttt{ShakeNDry}. The deformation part in \texttt{Crash} and the body movement in \texttt{Kobe} also verify this point. STFormer produces good visual results with accurately predicted motions in challenging cases. However, due to a high compression ratio and a lack of intermediate events, the overall visual quality is inferior to our method which produces sharper details and accurate motions. 

\subsubsection{Complexity analysis}
We present the number of model parameters and FLOPs of different algorithms in the right panel of Table~\ref{tab:res_gray}. Here the evaluation of FLOPs is based on outputting 16 gray-scale frames with size 256$\times$256 pixels. 
For STFormer+VFI, we sum up the number of parameters and FLOPs of individual models. 
VFI's FLOPs are calculated by summing up the FLOPs of interpolating the total number of frames. 
The comparison shows that the VFIformer and STFormer models possess fewer parameters while the Time Lens model contains many more, and an opposite trend of FLOPs is noticed. This might be attributed to the fact that VFIformer and STFormer are Transformer-based architectures that are of high computational complexity while Time Lens is CNN-based with relatively low computational demand. 
Our proposed network is composed of both Transformer and convolutional blocks, thus the number of parameters as well as FLOPs lie in between two groups of methods.

\begin{figure*}[]
    \centering
    \includegraphics[width=1\linewidth]{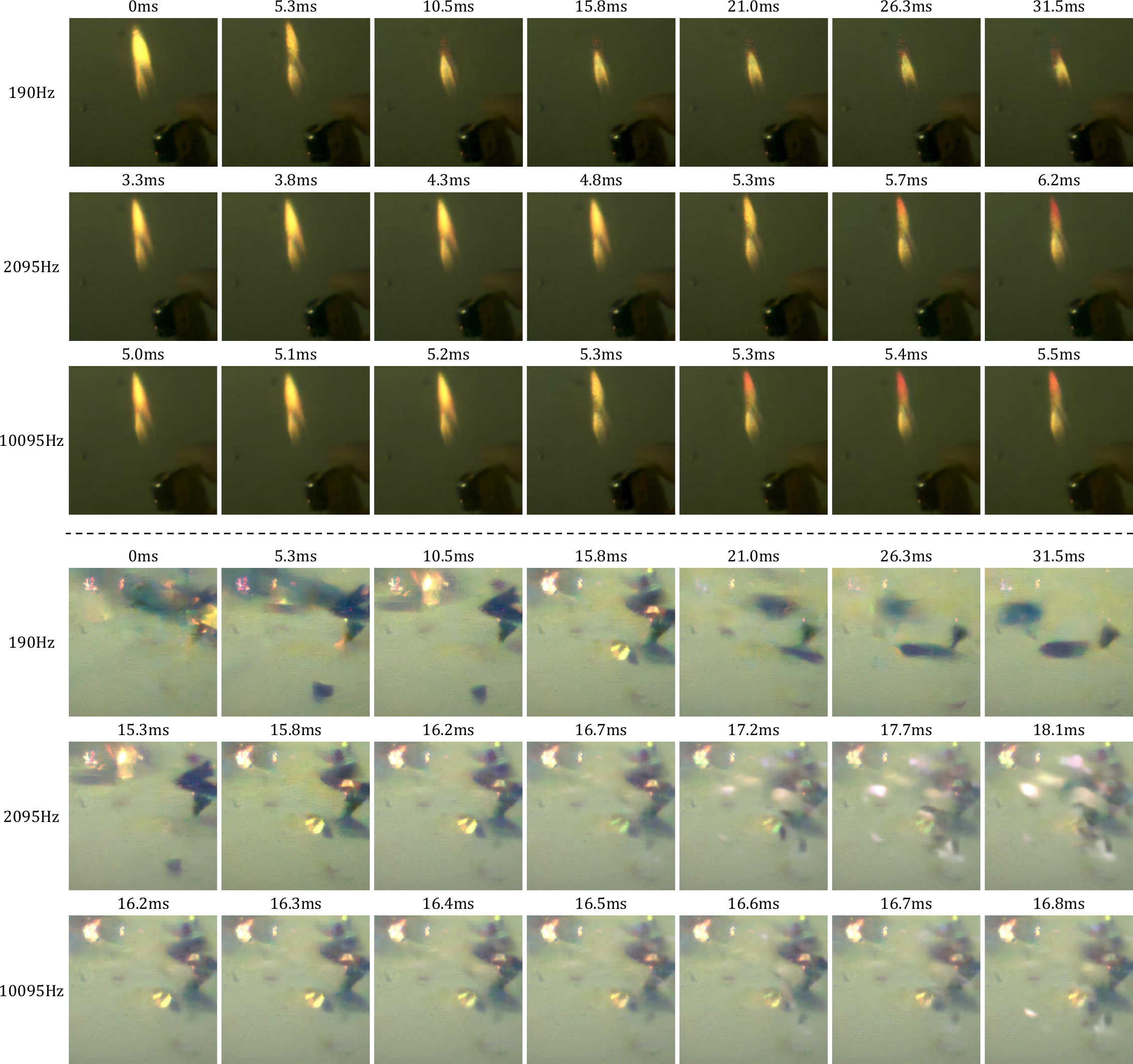}
    \caption{Videos of two challenging fast dynamics captured by our setup---extinguishing of a lighter flame (upper) and a burst of fireworks (bottom). In each scene, we display the down-sampled frames from the reconstructed 10K-FPS video at increasingly denser intervals from top to bottom, corresponding to frame rates at 190Hz, 2095hz, and 10095hz respectively. 
    }
    \label{fig:res_real}
    \vspace{-2mm}
\end{figure*}


\subsection{Experiments on real data}
To demonstrate the performance and applicability of the proposed imaging scheme in real scenarios, We use the built prototype shown in Fig.~\ref{fig:optics} (b) to capture high-speed scenes.
\subsubsection{Experimental settings}
For each scene, we set the CMOS's working frame rate to 24 FPS, encode 8 frames within one intensity snapshot, and record the events continuously with the synchronized event camera. Then the outputs from two sensors are preprocessed and fed into the reconstruction algorithm to retrieve the frame sequence. 

\vspace{1mm}
\noindent\textbf{Pre-processing.~~~~}
After obtaining raw coded measurements and events, we implement a standard pre-processing procedure. 
In the first place, we calibrate the masks and normalize the coded measurements following the steps in~\cite{Zhang:22}. 
In the second place, a registration algorithm is implemented to align the event voxels from DVS and coded measurements recorded by CMOS.   

The registration between the two modes is non-trivial and some implementation details in the above steps are crucial for the registration accuracy. The detailed steps are as follows: 
Firstly, we accumulate the events within the exposure elapse of the snapshot into an event image. 
Then, we crop the intensity/event patches from the full-size images and resize the event image to match the intensity image's resolution. The reason for patch-wise registration lies in two facts: (i) two arms are of different geometric aberrations, so global registration is difficult for a small number of registration parameters and suffers from low registration precision; (ii) due to GPU memory limit, the reconstruction network can only handle patch-wise data for reconstruction, which is in harmony with patch-wise registration.
Next, we use a similarity-based algorithm to align the event image patch toward the intensity counterpart and save the registration parameters. Here we choose to warp event signals toward the intensity measurement for two reasons: Event voxels are in float precision, so resizing, interpolation and alignment will not sacrifice precision; snapshot-coded measurement differs from a natural image, and the processing might involve sub-pixel interpolation and cause mismatch between the coded snapshot with the calibrated mask.
Finally, we use the registration parameters to align the cropped event voxels.


\vspace{1mm}
\noindent\textbf{Reconstruction.~~~~}
After calibration and pre-processing, the registered event voxels and intensity measurement are fed into the proposed dual-branch Transformer for frame reconstruction. To avoid the domain gap between simulated and real data, we use the model trained on the BS-ERGB dataset with real-captured videos and corresponding events to reconstruct the dense video frames. Afterward, we fine-tune the model for one epoch with the pre-calibrated masks, measurement, and aligned event representations of our setup, to further reduce the potential performance degeneration due to the difference between the simulated masks and the real-captured ones.

\subsubsection{Reconstructed results}
Following the above steps, we capture and reconstruct video frames of challenging scenes with high dynamics. Two representative visual results are displayed in Fig.~\ref{fig:res_real}, the extinguishing process of a flame when we turned off a lighter and a burst of fireworks. 
The final achieved frame rate is 10095Hz, and we display three sub-videos at increasing sampling rates from top to bottom to ``zoom in" the video along the temporal dimension, corresponding to videography at 190Hz, 2095Hz, and 1009 Hz respectively. 
It can be seen that the proposed method can reconstruct the fast dynamics when the flame gradually goes out and the fireworks explode, containing the intermediate dynamics between consecutive frames of conventional SCI. 
The subtle variations happening at a 0.1ms interval demonstrate the demanding imaging speed in studying the physical processes and the capability of our event-enhanced snapshot compressive imaging scheme. At such a high frame rate with mega-pixel resolution, the low bandwidth and cost additionally offer a great advantage over commercial high-speed cameras with similar performance specifications. 


\section{Conclusion and Discussions}\label{sec:conclusion}
In this paper, we present a new computational ultra-fast imaging mechanism featuring low-bandwidth data capture and ultrahigh-frame-rate video reconstruction. 
The proposed scheme successfully leverages the fast response of event cameras and enhances the SCI's capability of dynamic sensing by more than 50 times, resulting in 10k-frame reconstruction from a single snapshot. 
We experimentally validated the grand superiority of our method 
and demonstrated the feasibility in real applications as well.

The final frame rate is mainly limited by the event camera used in our implementation which has hundreds-of-millisecond resolution, in agreement with the 10K FPS reconstructed results. By using an event camera with higher temporal resolution, our proposed method is expected to achieve a higher reconstruction frame rate. 
In addition, despite the high performance, the setup is currently still a bench-top prototype and the reconstruction cannot work in real-time, which limits its applications on low-capacity platforms or scenarios lacking sufficient computing resources, such as drones and handheld cameras. In the future, lightweight setup and reconstruction will be developed to enable deployment on mobile platforms.

\section*{Acknowledgment}
This work is funded by National Natural Science Foundation of China (Grant No. 61931012 and 62088102).

\bibliographystyle{IEEEtran}
\bibliography{ref}

\newpage

\vfill

\end{document}